\documentclass[twocolumn]{article}

\usepackage{graphicx}
\usepackage{amsmath}
\usepackage[super,comma,sort&compress]{natbib} 
\usepackage{anysize}

\marginsize{2cm}{1cm}{1cm}{3cm}

\title{On deflection fields, weak-focusing and strong-focusing storage rings for polar molecules.}
\author{Adrian J. de Nijs and Hendrick L. Bethlem,\\ \small{Institute for Lasers, Life and Biophotonics, VU University Amsterdam} \\ \small{de Boelelaan 1081, 1081 HV Amsterdam, The Netherlands.}}
\date{}

\begin{document}

\twocolumn[\maketitle
 
\begin{abstract}In this paper, we analyze electric deflection fields for polar molecules in terms of a multipole expansion and derive a simple but rather insightful expression for the force on the molecules. Ideally, a deflection field exerts a strong, constant force in one direction, while the force in the other directions is zero. We show how, by a proper choice of the expansion coefficients, this ideal can be best approximated. We present a design for a practical electrode geometry based on this analysis. By bending such a deflection field into a circle, a simple storage ring can be created; the direct analog of a weak-focusing cyclotron for charged particles. We show that for realistic parameters a weak-focusing ring is only stable for molecules with a very low velocity. A strong-focusing (alternating-gradient) storage ring can be created by arranging many straight deflection fields in a circle and by alternating the sign of the hexapole term between adjacent deflection fields. The acceptance of this ring is numerically calculated for realistic parameters. Such a storage might prove useful in experiments looking for an EDM of elementary particles.\\ \\ 
\end{abstract}
]

\section{Introduction}

A neutral polar molecule in an inhomogeneous electric field experiences a force that is equal to its dipole moment times the gradient of the electric field strength. This force makes it possible to manipulate polar molecules using inhomogeneous electric fields in much the same way as that charged particles are manipulated using electric fields \citep{vandeMeerakker:2008}. In contrast to the forces on charged particles, however, the forces on polar molecules do not obey strict symmetries; i.e., for charged particles $\vec{\nabla} \vec{F} = 0$ everywhere in space, whereas for polar molecules this is only true in special cases\citep{Auerbach:1966}. As a consequence, manipulation tools for polar molecules suffer from aggravating non-linearities \citep{Lambertson:2003}.

One of the simplest manipulation tools is the electric or magnetic\footnote[2]{In this paper, we restrict ourselves to electric deflection fields, but this analysis also holds for magnetic deflection fields.} deflection field, dating back to the seminal experiments by Stern and Gerlach \citep{Friedrich:2003} in the 1930s. Deflection fields are extensively used to determine the magnetic and electric properties of atoms, molecules and clusters \citep{Miller:1988}. The ideal electric deflection field for polar molecules exerts a strong, constant force in one direction, while the force in the other directions is zero. For a molecule that has a linear Stark shift in the applied field, this implies that the electric field magnitude in one direction is linearly dependent on its position, while it is constant in the other directions. Unfortunately, such an electric field is not allowed by Maxwell's equations. In order to obtain accurate values for the polarizability of atoms and molecules, it is necessary to know both the electric field magnitude and its gradient. In the early experiments, the two wire field geometry was used, as this field and gradient are directly calculable from the geometry, thus avoiding difficult and tedious measurements of these quantities\citep{Ramsey:Book}. With the advent of numerical methods to calculate electric fields from arbitrary electrode geometries, an analytical expression of the electric field is no longer necessary. Hence, one may wonder if a more suitable deflection field can be created.

Recently, Stefanov et al.\citep{Stefanov:2008} discussed the optimal shape of a deflection field and presented a optimized design. Although the analysis of Stefanov et al. results in a near-ideal field, it  gives little insight into the underlying principles and limitations. In this paper, we analyze deflection fields in terms of a multipole expansion, following an approach similar to the one used in Kalnins et al.\citep{Kalnins:2002} and Bethlem et al.\citep{Bethlem:2006}. We find a simple expression for the resulting force, and show how it can be optimized by a suitable choice of the expansion coefficients.

Our motivation for this study stems from an experiment that is being planned at the VU University Amsterdam. In this experiment, 2-photon microwave transitions will be measured in a molecular beam of metastable CO molecules \citep{Bethlem:2009,deNijs:inpress}, with the ultimate goal to detect or limit a possible variation of the proton to electron mass ratio. The measured transitions are between a state which has a rather strong Stark shift and a state that has virtually no Stark shift. The deflection field will deflect the molecules in the initially populated state while it will not affect the molecules in the excitepd state. Hence, by using a position sensitive detector, the fraction of molecules that have made a transition can be recorded. In order to have sufficient signal to noise, we require a deflection field with a large aperture that gives a clear seperation between molecules in either state.

Our paper is organized as follows; in Sec.~\ref{Sec:DeflectionField}, we analyze deflection fields in terms of a multipole expansion and find a simple expression for the resulting force. We simulate the trajectories of metastable CO molecules through deflection fields with different expansion coefficients, and show how the expansion coefficients should be chosen. The deflection fields discussed in Sec.~\ref{Sec:DeflectionField} can be used to create weak-focusing and strong-focusing (alternating-gradient) storage rings for polar molecules. In Sec.~\ref{Sec:WeakStorageRing}, we show that unwanted terms in the deflection field limit the velocity of the molecules that can be stored in weak-focusing storage ring. In Sec.~\ref{Sec:StrongStorageRing}, we present a simple design for a strong-focusing storage ring and calculate its acceptance.

\section{Force on a polar molecule in a deflection field}
\label{Sec:DeflectionField}

In a region devoid of charges the electric field can be derived from the
electrostatic potential $\Phi$ using $\vec{E}=-\vec{\nabla} \Phi$. 
In two dimensions, $\Phi$ may be represented by a multipole expansion\citep{Kalnins:2002,Bethlem:2006} as:

\begin{equation}
\begin{split}
\Phi(x,y) = \Phi_{0} \left[
\sum_{n=1}^{\infty}\frac{a_{n}}{n}
\left(\frac{r}{r_{0}}\right)^{n}\cos(n\theta) \right.\\
 + \left. \sum_{n=1}^{\infty}\frac{b_{n}}{n}
\left(\frac{r}{r_{0}}\right)^{n}\sin(n\theta) \right].
\end{split}
\label{Eq:multipole}
\end{equation}

\begin{figure}[!t]
\centering
\includegraphics[width=8cm]{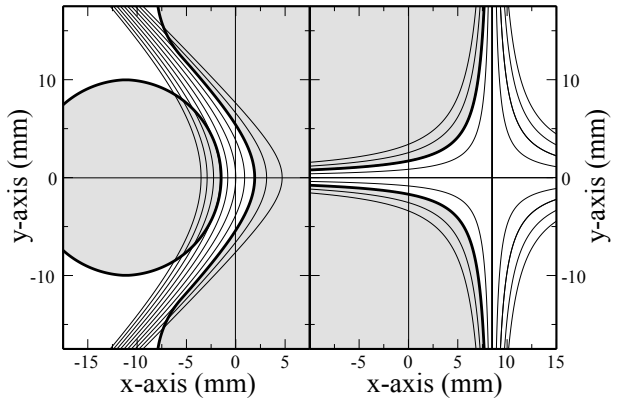}
\caption{Equipotential lines for two deflection fields with $\Phi_{0}=10$~kV,
$r_{0}=1.7$~mm, $a_{1}=1$ and $a_{2}=-0.2$ (left panel) and $\Phi_{0}=10$~kV,
$r_{0}=1.7$~mm, $b_{1}=1$ and $b_{2}=-0.2$ (right panel), all other coefficients are set to zero. The solid curves show the voltage in steps of 5~kV. At the center the voltage is 0. The fields can be created by placing an electrode at any of the potential lines. The bold curves shown in the left panel are the electrode surfaces chosen for our experiment at VU University Amsterdam. The origins of the graphs coincide with the molecular beam axis. 
}
\label{Fig:Phi}
\end{figure}

\noindent
Here $r=\sqrt{(x^{2} + y^{2})}$ and $\theta=\tan^{-1}\left(\frac{y}{x}\right)$ are the usual cylindrical coordinates. $a_{n}$ and $b_{n}$ are dimensionless constants. $r_0$ and $\Phi_{0}$ are scaling factors that characterize the size of the electrode structure and the applied voltages, respectively. The electric field magnitude at the centre is given by $E_{0} = (\Phi_{0}/r_{0})\sqrt{a_{1}^{2} + b_{1}^{2}}$. The $n=1$ terms in Eq.~\ref{Eq:multipole} represent a constant electric field, while the $n=2$ and $n=3$ terms represent the familiar quadrupole and hexapole fields that have been used extensively to focus molecules in low-field seeking states \citep{Reuss:Book:1988}.

Eq.~\ref{Eq:multipole} represents the most general form of the 2D electrostatic potential consistent with Laplace's equation. Now we choose suitable coefficients for making a deflection field. If we choose the molecules to be deflected in the horizontal plane, we can create a deflection field by setting all $b_{n}=0$, and setting $\vert a_{1} \vert \gg \vert a_{2} \vert \gg \vert a_{3} \vert$. This geometry is depicted on the left hand side of Fig.~\ref{Fig:Phi}. We will refer to this geometry as the 'AA' deflection field or 'conventional' deflection field. Alternatively, we may create a deflection field by setting all $a_{n}=0$, and setting $\vert b_{1} \vert \gg \vert b_{2} \vert \gg \vert b_{3} \vert$. This geometry is depicted on the right hand side of Fig.~\ref{Fig:Phi}. We will refer to this geometry as the 'BB' deflection field or wedge field. The solid curves in Fig.~\ref{Fig:Phi} show the voltage in steps of 5~kV. At the center the voltage is 0.  

\begin{figure}[!t]
\centering
\includegraphics[width=8cm]{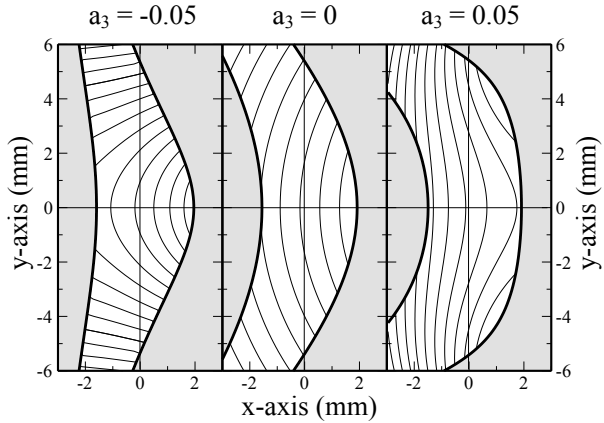}
\caption{The electric field magnitude for three different electrode geometries, with $\Phi_{0}=10$~kV, $r_{0}=1.7$~mm, $a_{1}=1$, $a_{2}=-0.2$ and $a_{3}$ as indicated in the figure. The solid curves show the magnitude of the electric field in steps of $5$~kV/cm. The electric field magnitude is $58.8$~kV/cm at the center and it decreases towards the right. }
  \label{Fig:Efields}
\end{figure}

These fields can be created by placing an electrode at any of the potential lines. The bold curves shown in the left panel are the electrode surfaces chosen for our experiment at VU University Amsterdam. The negative electrode is a cylinder with a radius of 10~mm, centered at $x=$-11.7~mm to which a voltage of $-10$~kV is applied. The positive electrode follows the contour $\Phi=10$~kV exactly up to $y=\pm12$~mm and is then rounded of with a radius of 10~mm. Because these electrodes do not match the equipotential exactly, higher order terms are introduced. A fit to the numerically calculated field shows that the coefficients are changed by less than 3\%.

Although the resulting potentials and electric fields for the AA and BB geometries are different, the magnitude of the electric field and forces are the same. We can write the electrostatic potential for the AA field as:

\begin{equation}
\Phi(x,y) =  \Phi_{0}\left(a_{1}\frac{x}{r_{0}} +
a_{2}\frac{\left(x^{2} - y^{2}\right)}{2r^{2}_{0}} + 
a_{3}\frac{\left(x^{3} -3xy^{2}\right)}{3r^{3}_{0}} \right).
\label{Eq:Phi}
\end{equation} 

\noindent 
From this potential, we can obtain the electric field magnitude, via: 

\begin{equation}
E(x,y)=\sqrt{\left(\frac{\partial\Phi}{\partial x}\right)^{2}
+\left(\frac{\partial\Phi}{\partial y}\right)^{2}}.
\label{Eq:E}
\end{equation}

\begin{figure}[!t]
\centering
\includegraphics[width=8cm]{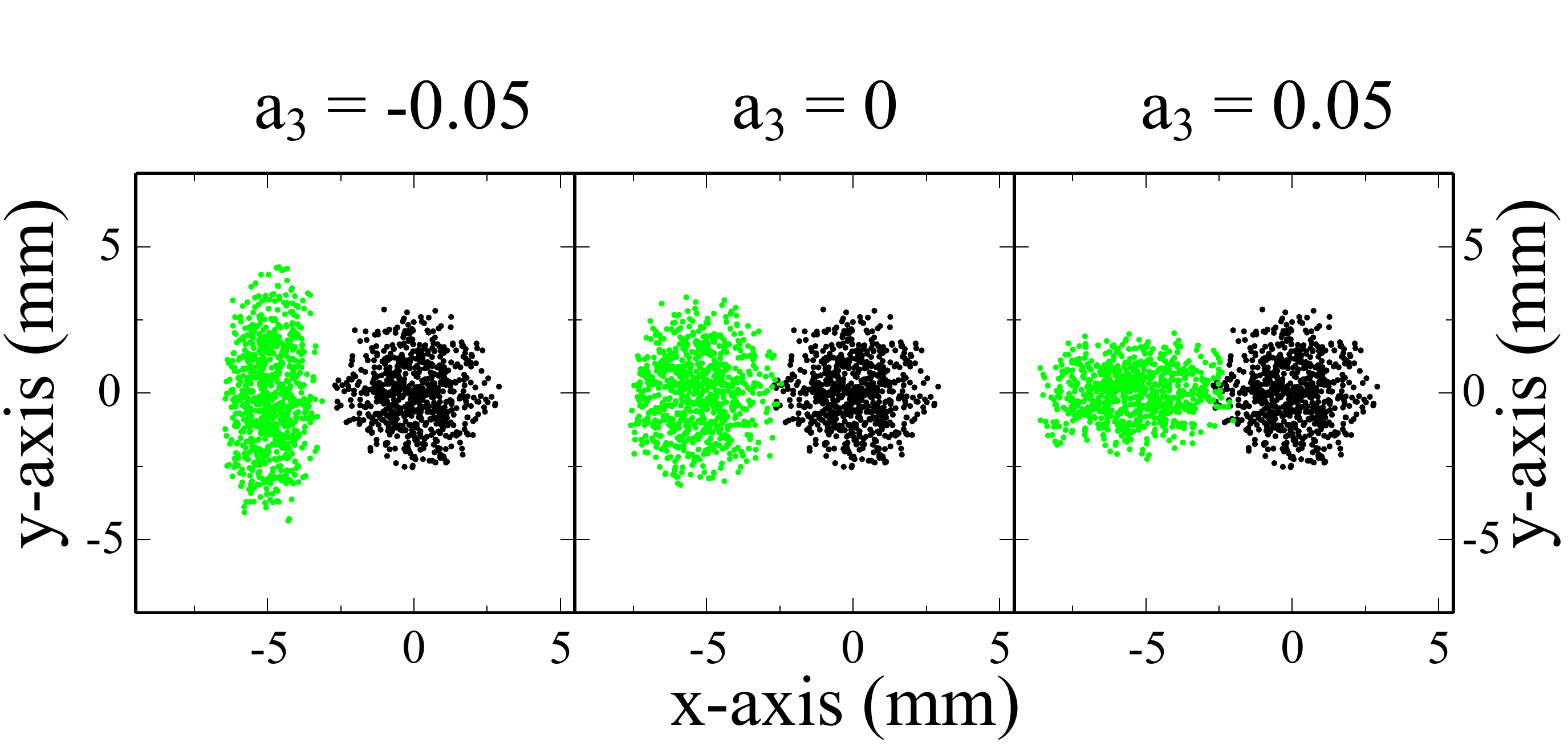}
\caption{Simulated transverse distribution of a beam of CO (a$^{3}\Pi$) molecules in the high-field seeking $J=6, M\Omega=6, \Omega=1$ state after passing through the deflection fields shown in Fig.~\ref{Fig:Efields}. In each panel, the distribution is shown with the field switched on (green dots) and off (black dots). When the fields are on, the beam is displaced from the molecular beam axis by about ~-5~mm over 1~meter of free flight.
}
  \label{Fig:deflection}
\end{figure}

In Fig.~\ref{Fig:Efields}, the electric field magnitude is shown for three different electrode geometries, with $\Phi_{0}=10$~kV, $r_{0}=1.7$~mm, $a_{1}=1$, $a_{2}=0.2$ and $a_{3}$ as indicated in the figure. The solid curves show the magnitude of the electric field in steps of $5$~kV/cm. The electric field magnitude is $58.8$~kV/cm at the center and it decreases towards the right.

From the electric field magnitude, we can obtain the Stark shift and the force on the molecules,via:

\begin{equation}
\vec{F}\left(\vec{r}\right) = -\vec{\nabla}  W(E)
= - \mu_\mathrm{eff} \vec{\nabla} E.
\label{Eq:dipoleforce}
\end{equation}

\noindent
Here we assume that the molecules experience a linear Stark shift; $W=-\mu_{\mathrm{eff}}(E_{0}) E$, in the applied field, with $\mu_{\mathrm{eff}}(E_{0})$ being the effective dipole moment of the molecule in the electric field at the center of the deflector. $\mu_{\mathrm{eff}}$ is positive for molecules in high field seeking states and negative for molecules in low field seeking states. Note that the Stark shift is assumed to be linear only over a small range of electric fields, this is a rather good approximation even for molecules that have a quadratic Stark shift. Throughout the region $r < r_{0}$, we can expand the force resulting from Eq.~\ref{Eq:Phi} as:

\begin{equation}
\begin{split}
F_{\mathrm{Stark},x} = \mu_{\mathrm{eff}}E_{0}\left[
\frac{a_{2}}{a_{1}}\frac{1}{r_{0}} +  2\frac{a_{3}}{a_{1}} \frac{x}{r_{0}^2}\right.\\
\left.- \left(\frac{1}{2}\left(\frac{a_{2}}{a_{1}}\right)^3 - 2 \frac{a_{2}}{a_{1}}\frac{a_{3}}{a_{1}}\right)
\frac{y^2}{r_{0}^3} + \ldots  \right],
\end{split}
\label{Eq:forcex}
\end{equation}

\begin{equation}
\begin{split}
F_{\mathrm{Stark},y} = \mu_{\mathrm{eff}}E_{0}\left[
\left(  \left( \frac{a_{2}}{a_{1}}\right)^2 -2\frac{a_{3}}{a_{1}} \right) \frac{y}{r_{0}^2}\right.\\
\left. - \left(\left(\frac{a_{2}}{a_{1}}\right) ^3 - 4 \frac{a_{2}}{a_{1}}\frac{a_{3}}{a_{1}}  \right) 
\frac{xy}{r_{0}^3} + \ldots \right].
\end{split}
\label{Eq:forcey}
\end{equation}

\noindent
Ideally, the deflection force is constant and strong in the $x$-direction, while it is zero along the $y$-direction. Thus, we would like to keep only the first term of Eq.~\ref{Eq:forcex} and set all other terms in Eq.~\ref{Eq:forcex} and Eq.~\ref{Eq:forcey} equal to zero. We see that the desired term scales as $a_{2}/a_{1}$ while the undesired terms scale as $a_{3}/a_{1}$ or as the second or third power of $a_{2}/a_{1}$. Thus, the undesired terms can be made arbitrary small, by choosing  $a_{3}=0$ and $a_{2}/a_{1} \ll 1$, but at the expense of the strength of the deflection force. In practice, one usually cannot afford to choose $a_{2}/a_{1}$, much smaller than $1/5$. The dominant undesired term in this case is the first term of Eq.~\ref{Eq:forcey}. This term can be cancelled with an appropriate choice of $a_{3}$, but this introduces other unwanted terms.

In order to study the influence of $a_{3}$ we have performed simulations of the trajectories of polar molecules though different deflection fields. In our simulations, we assume that a beam of molecules traveling with a forward velocity of $800$~m/s passes two diaphragms with a diameter of $1$~mm, spaced $50$~cm from each other before entering a $30$~cm long deflection field. After passing the deflection field, the molecules travel 100~cm further before being detected on a position sensitive detector. The trajectories within the deflector are calculated by numerical integration of the force derived from Eq.~\ref{Eq:forcex} and Eq.~\ref{Eq:forcey} (including all higher order terms) using a Runge-Kutta method. In these calculations, $\mu_{\mathrm{eff}}$ is taken to be 0.2~D, corresponding to 0.33$\times10^{-2}~$cm$^{-1}$/(kV/cm), which is the effective dipole moment of metastable CO in the $J=6$, $M\Omega=6$, $\Omega=1$ in an electric field of 58.8~kV/cm. If no deflection field is used, the transverse distribution of the beam at the detector is perfectly symmetric and has a FWHM of 2.6~mm. The arrival position of the undeflected molecules are shown as the black dots in Fig.~\ref{Fig:deflection}, whereas the arrival position of the deflected molecules are shown as the green dots. The three panels show the distributions when the three electrode geometries shown in Fig.~\ref{Fig:Efields} are used. The beams are deflected over an angle given by:

\begin{equation}
\theta = \mathrm{tan}^{-1} \left[\frac{\mu}{m}E_{0} \left(\frac{a_{2}}{a_{1}}\right) \frac{L_{\mathrm{defl}}}{r_{0} } \frac{1}{v_{z}^2}\right],
\label{Eq:deflectionangle}
\end{equation}

\noindent 
with $L_{\mathrm{defl}}$ being the length of the deflection field. In our case this corresponds to an angle of $\approx -4.5$~mrad which translates into a displacement from the molecular beam axis of $-5.2$~mm after 1~m flight distance (note that the beam is already displaced by 0.7~mm from the molecular beam axis at the exit of the deflector). From Fig.~\ref{Fig:deflection}, it is seen that the shape of the beam is deformed by the unwanted terms present in the force field. When $a_{3}=0$, the dominant term is the term linear in $y$, which causes the beam to be defocused in the $y$ direction. This is shown in the middle panel of Fig.~\ref{Fig:deflection}. When $a_3$ is large and negative, the beam is focused in the $x$-direction and defocused in the $y$-direction, this is shown in the left panel of Fig.~\ref{Fig:deflection}. When $a_3$ is large and positive the beam is focused in the $y$-direction and defocused in the $x$-direction, this is shown in the right panel of Fig.~\ref{Fig:deflection}.

\begin{figure}
\centering
\includegraphics[width=8cm]{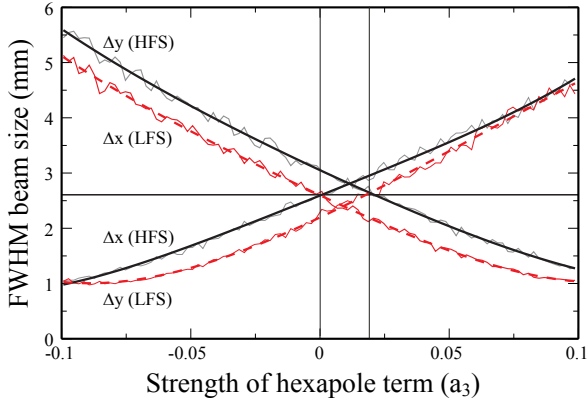}
\caption{FWHM of the transverse distribution of a beam of CO (a$^{3}\Pi$) molecules in the high-field seeking (HFS) $J=6, M\Omega=6, \Omega=1$ and low-field seeking (LFS) $J=6, M\Omega=-6, \Omega=1$ states as a function of the strength of the hexapole term $a_3$. The horizontal line shows the FWHM of the undeflected beam.} 
\label{Fig:FWHM}
\end{figure}

In Fig.~\ref{Fig:FWHM} the FWHM of the transverse distribution of a beam of CO (a$^{3}\Pi$) molecules in the high-field seeking (HFS) $J=6, M\Omega=6, \Omega=1$ and low-field seeking (LFS) $J=6, M\Omega=-6, \Omega=1$ states is shown as a function of  $a_3$. Let us first turn to the distribution of the high-field seekers, shown as the solid lines in the figure. It is seen that the beam is focused in the $x$-direction when $a_{3}/a_{1}<0$ and defocused when $a_{3}/a_{1}>0$. In the $y$ direction, the beam is focused when $a_{3}/a_{1}>0.5(a_{2}/a_{1})^{2}=0.02$ and defocused when $a_{3}/a_{1}<0.5(a_{2}/a_{1})^{2}=0.02$. The cross-section of the beam, proportional to the product of the two curves, is larger than that of the undeflected beam. Note that for molecules in low-field seeking states (shown as the dashed lines in Fig.~\ref{Fig:FWHM}) the situation is reversed; the cross-section of the beam is smaller, irrespectively of $a_{3}$. And, when $0<a_{3}/a_{1}<0.5(a_{2}/a_{1})^{2}$, low-field seekers are focused in both directions. 

The optimal choice of $a_{3}$ depends to some degree on experimental details. If one works with high-field seekers only, one might choose $a_{3}$ to be slightly negative such that the separation between molecules in different states becomes larger. In our experiment on metastable CO, we will use both high-field seekers and low-field seekers and we chose $a_{3}$ to be close to zero. It is noted that in molecular beam deflection experiments that use a two-wire field, often the 
the molecular beam is chosen to be at the position where the field is most homogeneous in the $y$-axis. This is referred to as the 'Hamburg' geometry \citep{Ramsey:Book} and corresponds to choosing  $a_{3}/a_{1} = 0.5(a_{2}/a_{1})^{2}$. 

\section{A weak-focusing storage ring for polar molecules.}
\label{Sec:WeakStorageRing}

A simple storage ring for polar molecules in high-field seeking states can be formed by bending the deflection fields discussed in the previous session into a circle. Such a storage ring is the direct analogue of the weak-focusing cyclotron for charged particles demonstrated by Lawrence in 1933 \citep{Lawrence:1931}. In a cyclotron, a magnetic field is used to bend charged particles in a circle while they are accelerated using electric fields. It was shown by Bethe and Rose \citep{Bethe:1937} and Kerst and Serber \citep{Kerst:1941} that the trajectories in a cyclotron are only stable if the magnetic field drops off slightly as a function of its radius. In this section, we will calculate the required form of the electric field of a weak-focusing storage ring for polar molecules. 

Let us assume that a BB deflection field similar to the one shown in Fig.~\ref{Fig:Phi} is bent into a circle of radius, $R_{\mathrm{ring}}$ (we chose a BB deflection field as it seems easier to inject and detect molecules in such a geometry). The $y$-direction is chosen to be the vertical direction, while $x$ is chosen to be in the plane of the ring, such that $x=r-R_{\mathrm{ring}}$. In order for the trajectories to be stable, two conditions must be met; (i) the applied force must vanish at the equilibrium orbit (chosen to be at $x=0, y=0$) and, (ii) for small displacements, the force should tend to restore the particle towards the equilibrium orbit. In keeping with literature on cyclotrons, we introduce the so-called field index, $n$, via:

\begin{equation}
n(r) =- \frac{\partial F_{\mathrm{Stark}}/F_{\mathrm{Stark}}}{\partial r/r}
= - \frac{r}{F_{\mathrm{Stark}}}\frac{\partial F_{\mathrm{Stark}}}{\partial r}.
\label{Eq:fieldindexgen}
\end{equation}

\noindent 
For $r=R_{\mathrm{ring}}$ and  $F_{\mathrm{Stark}}=\sqrt{F_{\mathrm{Stark},x}^2+F_{\mathrm{Stark},y}^2}$ given by Eq.~\ref{Eq:forcex} and Eq.~\ref{Eq:forcey}, this leads to:

\begin{equation}
n(r=R_{\mathrm{ring}})=- 2\frac{b_3}{b_2}\frac{R_{\mathrm{ring}}}{r_0}.
\label{Eq:fieldindex}
\end{equation}

\noindent 
If we neglect terms that are non-linear in the position, the force in the $x$-direction is given by:

\begin{equation}
\begin{split}
F_x&=  F_{\mathrm{centrifugal}} + F_{\mathrm{Stark},x}  \\
&= \frac{mv_{\varphi}^2}{R_{\mathrm{ring}}+x}+\mu_{\mathrm{eff}}E_{0}\left(
\frac{b_{2}}{b_{1}}\frac{1}{r_{0}} +  2\frac{b_{3}}{b_{1}} \frac{x}{r_{0}^2}\right) \\
&= \frac{mv_{\varphi}^2}{R_{\mathrm{ring}}}\left(1-\frac{x}{R_{\mathrm{ring}}}\right)
+\mu_{\mathrm{eff}}E_{0}\left(
\frac{b_{2}}{b_{1}}\frac{1}{r_{0}} +  2\frac{b_{3}}{b_{1}} \frac{x}{r_{0}^2}\right)\\
&= \left[ \frac{mv_{\varphi}^2}{R_{\mathrm{ring}}}
+\mu_{\mathrm{eff}}E_{0}\frac{b_{2}}{b_{1}}\frac{1}{r_{0}}\right]
-\left[\frac{mv_{\varphi}^2}{R_{\mathrm{ring}}^2}-2\mu_{\mathrm{eff}}E_{0}\frac{b_{3}}{b_{1}} \frac{1}{r_{0}^2}\right]x,
\end{split}
\end{equation}

\noindent 
with $v_{\varphi}$ the longitudinal velocity. The first stability condition requires $F_{x} =0$ at the equilibrium orbit. This leads to:

\begin{equation}
\frac{b_{2}}{b_{1}} =  -\frac{mv_{\varphi}^2}{\mu_{\mathrm{eff}}E_{0}}  \frac{ r_{0}}{R_{\mathrm{ring}}}.
\label{Eq:equilibrium}
\end{equation}

\noindent 
For the second stability condition we write $F_x$, as:

\begin{equation}
F_x \equiv -k_{x}x=-\left[\frac{mv_{\varphi}^2}{R_{\mathrm{ring}}^2}-2\mu_{\mathrm{eff}}E_{0}\frac{b_{3}}{b_{1}} \frac{1}{r_{0}^2}\right]x.
\label{Eq:Fxfinal}
\end{equation}

\noindent
Thus, molecules will oscillate around the equilibrium axis with an angular frequency given by:

\begin{equation}
\begin{split}
\omega_x &=\sqrt{\frac{k_x}{m}}=\sqrt{\frac{v_{\varphi}^2}{R_{\mathrm{ring}}^2}-2\frac{\mu_{\mathrm{eff}}E_0}{m}\frac{b_3}{b_1}\frac{1}{r_{0}^2}}\\
&=\Omega\sqrt{1-n}.
\end{split}
\label{Eq:omegax}
\end{equation}

\noindent
with $\Omega= v_{\varphi}/ R_{\mathrm{ring}}$ being the cyclotron frequency, and $n$ as defined in Eq.~\ref{Eq:fieldindex}.

Similarly, the force in the $y$-direction can be written as:

\begin{equation}
\begin{split}
F_{y} \equiv -k_{y}y =\mu_{\mathrm{eff}}E_{0}\left(
\left(  \left( \frac{b_{2}}{b_{1}}\right)^2 -2\frac{b_{3}}{b_{1}} \right) \frac{y}{r_{0}^2}\right). 
\end{split}
\label{Eq:ky}
\end{equation}

\noindent
And thus:

\begin{equation}
\begin{split}
\omega_y&=\sqrt{\frac{k_{y}}{m}}=
\sqrt{-\frac{\mu_{\mathrm{eff}}E_{0}}{m}
\left(  \left( \frac{b_{2}}{b_{1}}\right)^2 -2\frac{b_{3}}{b_{1}} \right) \frac{1}{r_{0}^2}}\\
&= \Omega\sqrt{n- \frac{mv_{\varphi}^2}{\mu_{\mathrm{eff}}E_{0}}}.
\end{split}
\label{Eq:omegay}
\end{equation}

\noindent 
In order to have stable confinement, we require the oscillation frequencies in both directions to be real. This is the case when:

\begin{equation}
\frac{mv^2_{\varphi}}{\mu_{\mathrm{eff}}E_{0}} < n < 1.
\label{Eq:stabilitycyclotron}
\end{equation}

\noindent 
Thus, for high-field seekers, stability is only possible when the Stark shift of the molecule is larger than two times the kinetic energy of the molecules, regardless of $n$. Hence a weak-focusing storage ring for polar molecules in high-field seeking states can only store beams at very low velocity. For instance, for CO (a$^3\Pi$) molecules in the $J=1, M\Omega =1, \Omega=1$ state in an electric field of 58.8~kV/cm, the maximum velocity that can be stored is about 15~m/s. It should be noted that in this derivation we have used $R_{\mathrm{ring}} \gg {r_0}$. For smaller rings, the multipole expansion should be written in cylindrical coordinates\citep{Nishimura:2004}. This decreases the stability region by a factor of $1/3$.

It is interesting to, once again, compare a weak-focusing storage ring for polar molecules in high-field seeking states with a cyclotron for charged particles. In a cyclotron, trajectories are stable when $0<n<1$. Thus, the analogy would be complete, if it wasn't for the left-hand side of  Eq.~\ref{Eq:stabilitycyclotron} being unequal to zero. This term arises from the fact that for polar molecules in electric fields,  $\vec{\nabla}\vec F$ is not necessarily equal to zero\citep{Auerbach:1966}. As a result of the extra term, the stability region for high-field seekers is decreased, while the stability region for low-field seekers is increased. Consequently, it is easy to construct a stable weak-focusing storage ring for low-field seekers \citep{Crompvoets:2001} but near impossible to store high-field seekers in a weak-focusing ring.

As a final note, we consider the so-called wire traps, proposed by Sekatskii\citep{Sekatskii:1995}, Sekatskii and Schmiedmayer\citep{Sekatskii:1996} and Jongma et al.\citep{Jongma:1997}. These are based on the fact that the electric field between two coaxial electrodes scales with the distance $r$ from the axis as 1/$r$. This implies that molecules with a linear Stark effect experience a force that scales as 1/$r^{2}$, and will be captured in stable "planetary" orbits. By comparison with the stability criterion, Eq.~\ref{Eq:stabilitycyclotron}, we see that the electric field in a wire trap drops too quickly ($n=$2) and the motion is unstable according to the used definition. Molecules in a wire trap are not stably confined around a certain equilibrium orbit, they are merely confined around the wire. Consequently, cooling techniques, such as sympathetic or evaporative cooling, cannot be applied to molecules in a wire trap (or to molecules in a storage ring based on a toroidal wire\citep{Loesch:2000}).

\section{AG ring}
\label{Sec:StrongStorageRing}

In a cyclotron for charged particles, the trajectories are stable in the horizontal and vertical plane if the field index, $n$, is between 0 and 1. In this case the oscillation frequencies are always a fraction of the cyclotron frequency. It was shown by Courant and Snyder in 1953 \citep{Courant:1958}, that much stronger confinement can be achieved by alternating the field index between a large positive and a large negative value. In this case, the particles are alternately focused and defocused in both planes. As the particles are, on average, further away from the equilibrium orbit when the field is focusing and closer to the equilibrium orbit when the field is defocusing, the trajectories are stable in both planes. These rings are called strong-focusing or alternating-gradient (AG) storage rings.

The application of AG focusing for polar molecules was first discussed by Auerbach et al.\citep{Auerbach:1966} and first demonstrated by Kakati and Lain\'{e}\citep{kakati:1967}. More recently AG focusing was used for decelerating\citep{Tarbutt:2004,Bethlem:2006} and guiding \citep{Filsinger:2008, Wall:2009} beams of heavy polar molecules. Nishimura et al.\citep{Nishimura:2004} presented a design for a three meter diameter AG storage ring
capable of storing beams with a forward velocity of 30~m/s. This AG-ring consists of eight octants, each one containing a bend element, a buncher and a pair of alternating-gradient focusing triplets. Here, we propose an AG ring consisting of 40 straight deflection fields arranged in a circle, similar to a storage ring for molecules in low-field seeking states that was recently demonstrated by Zieger et al.\citep{Zieger:2010}. Each deflection fields contains a strong quadrupole ($b_{2}$) term to deliver the necessary centripetal force and a hexapole ($b_{3}$) term to focus the molecules. By alternating the sign of the $b_{3}$ term in subsequent sections, molecules are alternately focused and defocused in both planes, leading to stable trajectories. 

In order to determine the stability of the ring, we use a computer code that was developed to simulate the trajectories through a storage ring for low-field seeking states (see Heiner\citep{Heiner:Thesis} for details).  In our calculation, we consider a storage ring with a radius, $R_{\mathrm{ring}}$, equal to 0.25~m, consisting of 40 straight segments with a length of about 39~mm. The segments are chosen to be of the BB-type shown on the right hand side of  Fig.~\ref{Fig:Phi}, with $\Phi_{0}=10$~kV, $r_{0}=1.7$~mm, $b_{1}=1$ and $b_{2}=-0.2$. The longitudinal velocity that can be confined in this ring follows from Eq.~\ref{Eq:equilibrium}. For CO~(a$^3\Pi$)  molecules in the high-field seeking $J=1, M\Omega=1, \Omega=1$ state with an effective dipole moment of 0.7D, corresponding to 0.011~cm$^{-1}$/(kV/cm),  the maximum velocity is 92~m/s. If the segments are indexed by $s$, the hexapole term in the $s^{\mathrm{th}}$ deflection fields is given by:

\begin{equation}
b_{3} = b_{3c} + (-1)^{s} b_{3a} = \frac{b_{2}}{4}\left(\frac{r_{0}}{R_{ring}}+\frac{b_{2}}{b_{1}}\right) +(-1)^{s} b_{3a},
\label{Eq:b3c}
\end{equation}

\noindent
such  that molecules are focused in the $x$ plane and defocused in the $y$ plane in odd segments, and vice versa in the even segments. The constant hexapole term, $b_{3c}$, is added to ensure that the working conditions in the $x$ and $y$ plane are similar. In the calculation, typically 10$^5$ molecules are generated with a random initial position and velocity. The acceptance is then found by multiplying the fraction of surviving molecules after a set number of roundtrips by the phase space volume of the initial packet. The aperture in the vertical direction is determined by the electrodes and in the horizontal direction by the non-linear terms in the force field.  

\begin{figure}[!t]
\centering
\includegraphics[width=8cm]{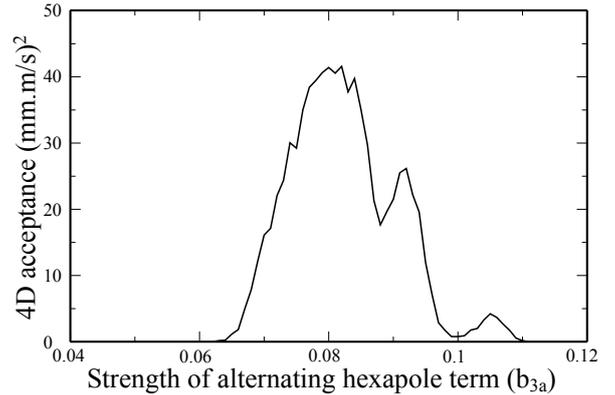}
\caption{Transverse acceptance of metastable CO molecules in a 0.5~meter diameter alternating-gradient ring consisting of 40 deflection fields as a function of  the alternating hexapole $b_{3v}$, with $\Phi_{0}=10$~kV, $r_{0}=1.7$~mm, $b_{1}=1$, $b_{2}=-0.2$ and $b_{3}= b_{3c} \pm b_{3a}$. The longitudinal velocity of the stored molecules is equal to 92~m/s.} 
\label{fig:AGring}
\end{figure}

In Fig.~\ref{fig:AGring} the transverse acceptance of the AG-ring is plotted as a function of  the alternating hexapole term, $b_{3a}$. We see that the acceptance peaks when  $b_{3a}$ is equal to about 0.08. This corresponds to the situation when the molecules make slightly less than halve an oscillation in a single segment, as expected from the theory on AG-focusing\citep{Bethlem:2006}. If the $b_{3a}$ is increased further the molecules are over-focused and the motion becomes unstable. At the optimal value of  $b_{3a}$ the acceptance is 40~(mm$\times$m/s)$^2$. This corresponds to a trap depth of about 8~mK and an effective aperture of about 2$\times$2~mm. The acceptance is about 200 times smaller than the acceptance calculated for the storage ring for low-field seeking states of  Zieger et al.\citep{Zieger:2010,Heiner:Thesis}, and about 4 times smaller than the acceptance calculated for an AG-guide for CaF molecules\citep{Wall:2009}. It should be noted, however, that the acceptance of AG focusing devices is very sensitive for the mechanical alignment of the AG-lenses, and numerical calculations usually greatly overestimate the acceptance \citep{Bethlem:2006}.

\section{Conclusion}
\label{Sec:Conclusion}

In order to make optimal use of manipulation tools for polar molecules it is important to understand their possibilities and limitations. In this paper, we analyze deflection fields in terms of a multipole expansion and find a simple expression for the resulting force. It is found that the field contains a term that focuses molecules in low-field seeking state and defocuses molecules in high-field seeking states perpendicular to the deflection direction. This term arises from the fact that  $\vec{\nabla} \vec{F}$ is unequal to zero. It can be made small by choosing $a_{2}/a_{1} \ll 1$, but this goes at the expense of the strength of the deflection force. The force in a two wire field in the 'Hamburg' geometry is homogeneous in the direction perpendicular to the deflection. This corresponds to a positive value for the hexapole term, a$_{3}$. From our simulations, we find that it is advantageous to choose a$_{3}$ to be close to zero or slightly negative. 

By bending a deflection fields into a circle a simple storage ring for polar molecules in high-field seeking states can be created. It is shown that a weak-focusing storage ring for polar molecules can only store beams at low velocities and is of little practical relevance. A strong-focusing (alternating-gradient) storage ring can be created by arranging many straight deflection fields in a circle and by alternating the sign of the hexapole term between adjacent deflection fields. The acceptance of such a ring is numerically calculated for realistic parameters and is found to be 40~(mm$\times$m/s)$^2$. Further study is necessary to optimize the geometry and to investigate how varying the applied fields to confine the molecules in the longitudinal directions affects the acceptance in the transverse direction. One application of such a storage ring might be for performing an EDM measurement. The sensitivity of EDM measurements scales with the time that the molecules spend in the interaction zone, hence various labs are working on ways to increase this time by decelerating and possibly trapping the molecules \citep{Tarbutt:2009}. It would seem possible to inject cryogenic beams\citep{Maxwell:2005,Patterson:2007} of heavy polar molecules such as YbF, PbF or WC, with a velocity below 200~m/s directly into a 2~m diameter storage ring.

\section{Acknowledgements}
\label{Sec:Acknowledgements}

This work is part of the research program of the "Stichting voor Fundamenteel Onderzoek der Materie
 (FOM)," which is financially supported by the "Nederlandse
 Organisatie voor Wetenschappelijk Onderzoek
 (NWO)." H. L. B. acknowledges financial support from NWO via a VIDI-grant, and from the ERC via a Starting Grant. We thank Mike Tarbutt, Peter Zieger and Wim Ubachs for useful discussions.

\footnotesize{\providecommand*{\mcitethebibliography}{\thebibliography}
\csname @ifundefined\endcsname{endmcitethebibliography}
{\let\endmcitethebibliography\endthebibliography}{}
}

\end{document}